\documentclass[pre,aps,twocolumn]{revtex4}
\usepackage{epsfig}
\usepackage{bm}

\newcommand{\C}[1]{{\mathcal{#1}}}
\newcommand{\pa}{\partial}
\usepackage[latin1]{inputenc}
\newcommand{\beq}{\begin{equation}}
\newcommand{\eeq}{\end{equation}}
\newcommand{\bea}{\begin{eqnarray}}
\newcommand{\eea}{\end{eqnarray}}
\begin{document}
\title{Elastic nonlinearities in a one-dimensional model of fracture}
\date{\today}
\author{Eran Bouchbinder$^{1}$ and Ting-Shek Lo$^{2}$}
\affiliation{$^1$Racah Institute of Physics, Hebrew University of Jerusalem, Jerusalem 91904, Israel,\\
$^2$Dept. of Chemical Physics, Weizmann Institute of Science, Rehovot 76100, Israel}
\begin{abstract}
The dynamics of rapid brittle cracks is commonly studied in the framework of linear elastic fracture mechanics where nonlinearities are neglected. However, recent experimental and theoretical work demonstrated explicitly the importance of elastic nonlinearities in fracture dynamics. We study two simple one-dimensional models of fracture in order to gain insights about the role of elastic nonlinearities and the implications of their exclusion in the common linear elastic approximation. In one model we consider the decohesion of a nonlinear elastic membrane from a substrate. In a second model we follow the philosophy of linear elastic fracture mechanics and study a linearized version of the nonlinear model. By analyzing the steady state solutions, the velocity-load relations and the response to perturbations of the two models we show that the linear approximation fails at finite crack tip velocities. We highlight certain features of the breakdown of the linear theory and discuss possible implications of our results to higher dimensional systems.
\end{abstract}
\maketitle

\section{Introduction}
\label{intro}

The dynamics of rapid brittle cracks exhibits a rich phenomenology that is not yet well understood. For example, crack tip instabilities (such as a side-branching instability \cite{99MF} and an oscillatory one \cite{07LBDF}), that were shown to govern fracture dynamics at high propagation velocities, are poorly understood from a fundamental point of view. The major stumbling block in developing a well-established theory of these phenomena is our lack of understanding of the physics of the ``fracture process zone'' within which nonlinear deformation, dissipation and material separation processes take place. The classic approach of linear elastic fracture mechanics (LEFM) assumes infinitesimal deformation outside this typically small process zone and predicts asymptotically ``diverging-like'' strain and stress fields \cite{98Fre}. Under these assumptions the energy flux into the process zone is calculated and an equation for the rate of crack growth is obtained by equating it to an unknown dissipation function $\Gamma$. This quantity lumps together all the poorly understood nonlinear and dissipative properties of the process zone dynamics \cite{98Fre}.

Very recent experimental and theoretical work demonstrated explicitly the existence of a nonlinear elastic zone in the near vicinity of a rapidly moving crack tip, where the deformation fields were shown to be quite different from those predicted by LEFM \cite{08LBF,08BLF}. One implication of these findings is that although LEFM may provide reasonable estimates of the energy flowing to the process zone, it fails to represent properly the ways in which breaking stresses are being transmitted to the crack tip. Therefore, as long as the path of the crack is known and is stable against perturbations, the energy based approach of LEFM seems useful; however, as such conditions are rarely met and the question of path stability is usually of prime importance, the near tip deformation and stress fields themselves may play a central role in describing fast fracture and the associated instabilities. Moreover, these tip instabilities seem to involve a non-geometrical lengthscale (for example, the minimal side-branch length \cite{96SF,05BP} or the wavelength of oscillations \cite{07LBDF,07BP}) that is missing in LEFM. Thus, the findings of \cite{08LBF,08BLF}, that suggest the existence of a dynamical lengthscale associated with the nonlinear elastic zone, shed new light on the search for a missing lengthscale.

Motivated by these recent results, we aim at gaining additional insights about the possible roles played by elastic nonlinearities in fracture dynamics and about the possible implications of neglecting these effects in the common linear elastic approach. For that aim we study in this paper simple one-dimensional models of fracture in which an elastic membrane is being detached from a substrate by the propagation of a decohesion front. The models are being defined by the following differential equation for the scalar deformation $u(x,t)$ of the membrane
\begin{equation}
\label{1d}
\rho\pa_{tt} u = \pa_x s-\kappa^2(u-\delta)-\phi(u,\pa_t u)\ ,
\end{equation}
where $\rho$ is the linear mass density of the membrane. The first term on the right-hand-side represents the force per unit length due to the deformation $u(x,t)$, where the stress $s(x,t)$ is related to $u(x,t)$ through a constitutive law of the form
\begin{equation}
\label{stress}
s=\mu {\C F}(\pa_x u) \ .
\end{equation}
Here $\mu$ is the elastic modulus of the membrane and the functional ${\C F}(\pa_x u)$ represents a general nonlinear stress-strain relation. The strain is the displacement gradient $\pa_x u$. The second term on the right-hand-side represents the loading of the membrane by elastic springs whose spring constant is $\kappa^2$ and whose natural length is reached when $u(x,t)\!=\!\delta$. Note that $\mu$ has the dimension of force, while $\kappa^2$ has the dimension of force/squared length. The term $\phi(u,\pa_t u)$ represents the visco-elastic interaction of the membrane with the substrate. This term should also include a criterion for detachment from the substrate, serving as a fracture criterion.

Similar one-dimensional models were studied previously in relation to various aspects of fracture dynamics, see for example \cite{98Fre,92Lan,93Mar,95CLN, 98LL}. Our model follows the spirit of these works and its linear approximation is similar to the model discussed in Ref. \cite{93Mar}. Our strategy is to study two related models with the same $\phi(u,\pa_t u)$, where in one we consider the solution for a nonlinear ${\C F}(\pa_x u)$ and in the other we follow the philosophy of LEFM and {\em linearize} ${\C F}(\pa_x u)$ first and then solve the model. Our main goal is to compare various aspects of the dynamics of the nonlinear and linearized models in order to gain insights about the role of elastic nonlinearities in the dynamics of fast brittle cracks.

Our results show that for sufficiently large $\kappa$ there exists a range of small crack tip velocities $v$ (i.e. $v\!\ll\!c$, where $c\!\equiv\!\sqrt{\mu/\rho}$ is the {\em linear} wave speed), such that the linear theory provides reasonable approximations to the nonlinear theory. However, as the velocity $v$ increases the linear approximation deteriorates progressively until it fails to capture important aspects of the dynamics. For smaller values of $\kappa$ there {\em exists no range of validity for the linear approximation}. Our conclusion is that in our simple one-dimensional model of fracture LEFM inevitably breaks down at finite crack tip velocities, in agreement with the findings of \cite{08LBF,08BLF}. This breakdown can be of significance to the understanding of crack tip instabilities in higher dimensional and more realistic fracture problems. Specifically, we show how elastic nonlinearities affect the limiting crack velocity, the strain and stress fields near the moving crack tip and the dynamical time and length scales involved in the physics of the near tip region.

In Sect. \ref{models} we present the models by first introducing a nonlinear constitutive law that can capture both softening and stiffening behaviors. Then we present a linearized version of this nonlinear model. In Sect. \ref{SS} we solve for the steady states of the models and discuss the resulting deformation profiles, the load-velocity curves and the limiting crack tip velocities. In Sect. \ref{stability} we study the response of the steady states to perturbations in the crack tip location within the framework of a linear stability analysis. In Sect. \ref{sum} we summarize the results and discuss their possible implications to more realistic, higher dimensional, systems.

\section{The models}
\label{models}

\begin{figure}
\centering
\epsfig{width=.45\textwidth,file=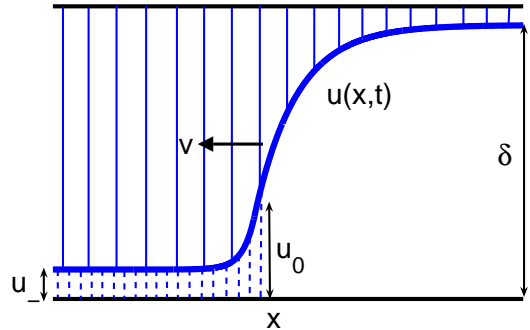}
\caption{(Color online) A sketch of the one-dimensional model. A membrane, whose profile is given by the deformation $u(x,t)$, propagates from right to left (in this example) at a velocity $v$. The vertical solid lines represent the loading springs. The vertical dashed lines represent the cohesive springs. $u_0$ is the breaking threshold of the cohesive springs, $\delta$ is the deformation of the membrane for which the loading springs are completely relaxed (see Eq. (\ref{1d})) and $u_-$ is the asymptotic value of the deformation in the negative x direction (see Eq. (\ref{u_neg})). }\label{sketch}
\end{figure}

In order to complete the definition of the models to be considered below, we should supplement Eq. (\ref{1d}) with a constitutive law ${\C F}(\pa_x u)$ in Eq. (\ref{stress}) and to specify the form of the interaction of the membrane with the substrate $\phi(u,\pa_t u)$. For the latter we choose
\begin{equation}
\label{phi}
\phi(u,\pa_t u)=\alpha^2uH(u_0-u)+\eta\pa_t uH(u_0-u)\ .
\end{equation}
This is the cohesion force that binds the membrane to the substrate and can be thought of as the action of visco-elastic springs with a spring constant $\alpha^2$, a small friction/viscosity-like coefficient $\eta$ and a breaking threshold $u_0$. The Heaviside step function $H(\cdot)$ represents the irreversible breaking of the cohesion springs when the displacement exceeds $u_0$. The small dissipation associated with $\eta$ is essential to ensure the existence of steady states, see \cite{93Mar} and below for details. A sketch of the one-dimensional model is shown in Fig. \ref{sketch}.
%%%%%%% FIGURE 1 %%%%%%%%%%%%%%%%%%
%%%%%%%%%%%%%%%%%%%%%%%%%%%%%%%%%%%

\begin{figure}
\centering
\epsfig{width=.45\textwidth,file=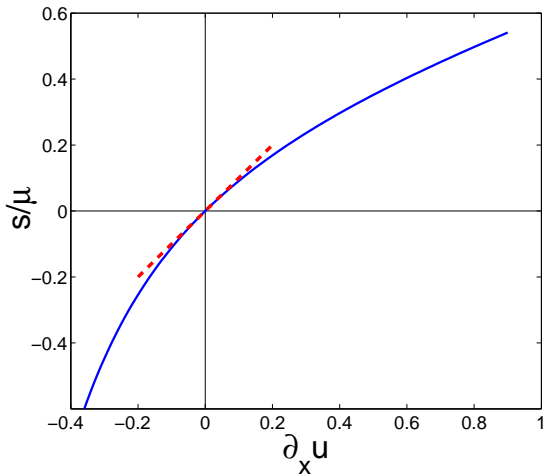}
\caption{(Color online) $\C F\!=\!s/\mu$ as a function of $\pa_x u$ (solid line). The linear approximation at small $\pa_x u$ is added (dashed line). The branch corresponding to $\pa_x u\!>\!0$ exhibits nonlinear softening, while the one that corresponds to $\pa_x u\!<\!0$ exhibits nonlinear stiffening.}\label{nonlin}
\end{figure}

The constitutive law for the membrane is chosen to be
\begin{equation}
\label{NH}
{\C F}(\pa_x u)=\frac{1}{3}\left[\left(1+\pa_x u\right)-\left(1+\pa_x u\right)^{-2} \right] \ .
\end{equation}
This law corresponds to the uni-axial behavior of the tensorial neo-Hookean constitutive law considered in \cite{08BLF}.
For small strains, Eq. (\ref{NH}) can be linearized, yielding
\begin{equation}
\label{LE}
{\C F}(\pa_x u)\simeq \pa_x u +{\C O}\left[(\pa_x u)^2 \right]\quad \hbox{for}\quad \pa_x u \ll 1 \ .
\end{equation}
The constitutive law of Eq. (\ref{NH}) is plotted in Fig. \ref{nonlin}. For positive strains, $\pa_x u\!>\!0$, the constitutive law exhibits nonlinear softening, i.e. the local tangent to the curve is smaller than the tangent at infinitesimal strains. For negative strains, $\pa_x u\!<\!0$, the constitutive law exhibits nonlinear stiffening, i.e. the local tangent to the curve is larger than the tangent at infinitesimal $\pa_x u$. Note that in the nonlinear softening case the tangent approaches $1/3$ at large strains. Usually, fracture is associated with positive strains. However, in our simple model the sign of $\pa_x u$ is determined by the direction of crack propagation; when the crack propagates from right to left we have $\pa_x u\!>\!0$ and the softening branch is selected, while when it propagates from left to right we have
$\pa_x u\!<\!0$ and the stiffening branch is selected. Thus, both nonlinear behaviors can be incorporated into our model and we refer to these cases as the nonlinear softening and nonlinear stiffening models respectively. When the linearized relation of Eq. (\ref{LE}) is used instead of Eq. (\ref{NH}), we refer to the model as the linear one. Note that in that case the direction of crack propagation is irrelevant as the linearized relation is symmetric.
%%%%%%% FIGURE 2 %%%%%%%%%%%%%%%%%%
%%%%%%%%%%%%%%%%%%%%%%%%%%%%%%%%%%%

In order to prepare Eq. (\ref{1d}) for the analysis to come we nondimensionalize all of the quantities by measuring length in units of $u_0$, velocity in units of $c$ and force in units of $\mu$. For simplicity we denote all the nondimensionalized quantities using their original notations. Furthermore, we set $\alpha\!=\!1$. We thus end with the following equation
\begin{equation}
\label{nondim}
\pa_{tt} u = \pa_x {\C F}(\pa_x u)-\kappa^2(u-\delta)-uH(1-u)-\eta \pa_t u H(1-u) \ ,
\end{equation}
where
\begin{equation}
\label{NNF}
\pa_x {\C F}(\pa_x u) = \frac{\pa_{xx} u}{3}\left[1+\frac{2}{\left(1+\pa_x u\right)^3} \right] \ ,
\end{equation}
for the nonlinear models and
\begin{equation}
\label{LF}
\pa_x {\C F}(\pa_x u) \simeq \pa_{xx} u \ ,
\end{equation}
for the linear one.

\section{Steady state solutions}
\label{SS}

Our goal in this section is to solve for the steady states of both the linear and nonlinear models and to compare various properties of these solutions. The tip of the crack is defined as the point where the cohesive springs break and assumed to propagate at a constant velocity $v$. In a co-moving frame $x'\!=\!x+vt$ the crack tip lies at the origin and $u(0)\!=\!1$. We now look for a steady state solution of the form $u(x+vt)$. We start by considering the linear model, for which one can obtain analytic results.

\subsection{The linear model}
\label{SSlin}

For steady state conditions, Eq. (\ref{nondim}), with Eq. (\ref{LF}), reduces to an ordinary differential equation of the form
\begin{equation}
\label{EOM_linearSS}
0= (1-v^2)\pa_{xx}u-\kappa^2(u-\delta)-(u+v\eta \pa_xu)H(-x)\ ,
\end{equation}
where we replaced $x'$ with $x$ for the simplicity of the notation.
The appearance of a step function, modeling the rupture of the cohesive springs at a critical displacement, implies that the differential equation is ill-defined at $x\!=\!0$. Therefore, we treat separately the positive and negative x domains and demand continuity of $u(0)$ and $\pa_xu(0)$. We first consider the domain $x\!>\!0$. In the limit $x\!\to\!+\infty$ we have $u\!\to\!\delta$. Therefore, we assume a solution of the form $u=\delta+(1-\delta)\exp{(kx)}$, that satisfies $u(0)\!=\!1$. Substituting in Eq. (\ref{EOM_linearSS}), we obtain a simple second order algebraic equation for $k$. We choose the negative root since $x\!>\!0$ (the solution must be bounded). Following a similar procedure for $x\!<\!0$, we arrive at
\begin{widetext}
\begin{eqnarray}
\label{solution_linearSS}
u(x)&=&\delta+(1-\delta)\exp{\left(-\frac{\kappa x}{\sqrt{1-v^2}}\right)},\hspace{6.28cm}\hbox{for}\quad x>0\ ,\nonumber\\
u(x)&=&\frac{\delta \kappa^2}{1+\kappa^2}+\left[1-\frac{\delta \kappa^2}{1+\kappa^2}\right]\exp{\left[\left(\frac{\eta v+\sqrt{\eta^2 v^2+4(1+\kappa^2)(1-v^2)}}{2(1-v^2)}\right)x\right]},\quad\hbox{for}\quad x<0\ .
\end{eqnarray}
\end{widetext}
where the velocity $v$ is still undetermined. The velocity is determined by demanding that $\pa_xu(0)$ is continuous. Note that higher order derivatives at $x\!=\!0$ are discontinuous due to the discontinuity of the force at this point. Using the continuity of the derivative at $x\!=\!0$ we obtain
\begin{equation}
\label{vSS}
v(\delta,\kappa,\eta)=\frac{\delta^2-\delta_c^2}{\sqrt{(\delta^2-\delta_c^2)^2+\kappa^2\eta^2(\delta-1)^2(\delta-\delta_c^2)^2}}\ ,
\end{equation}
with
\begin{equation}
\label{Griffith}
\delta_c\equiv \sqrt{1+\kappa^{-2}} \ .
\end{equation}

We are now able to interpret the steady state solutions of the linear model in terms of standard LEFM concepts. $\delta$, with a fixed $\kappa$, can represent the load and $\eta$ is a material parameter that is related to dissipation. Within this interpretation, Eq. (\ref{vSS}) tells us that $\delta_c$ is the critical load needed to initiate crack propagation. It is a direct analog of the Griffith criterion \cite{98Fre}. Furthermore, Eq. (\ref{vSS}) with a fixed $\kappa$ represents the velocity-load relation of the model. For $\delta$ close to $\delta_c$, the velocity of the crack is much smaller than the wave speed, while for sufficiently large $\delta$ the tip velocity approaches the wave speed. Thus, the limiting crack velocity is the wave speed, in complete analogy with the common prediction of LEFM in higher dimensions, where the limiting velocity is the Rayleigh wave speed \cite{98Fre}.

The role of the small dissipation coefficient $\eta$ is most clearly demonstrated by
multiplying Eq. (\ref{EOM_linearSS}) by $\pa_xu$ and integrating from $-\infty$ to $\infty$ to obtain
\begin{equation}
\label{energy_balance}
\frac{1}{2}(1-u_{-}^2)+\eta v \int_{-\infty}^0(\pa_xu)^2dx=\frac{1}{2}\kappa^2(\delta-u_{-})^2 \ ,
\end{equation}
where
\begin{equation}
\label{u_neg}
u_{-}\equiv\frac{\delta \kappa^2}{1+\kappa^2} \ ,
\end{equation}
is the value of $u$ as $x\!\to\!-\infty$.
It is important to note that this result holds for {\em any} elastic constitutive law ${\C F}(\pa_xu)$ and not only for a linear one.
The first term on the left-hand-side of Eq. (\ref{energy_balance}) is simply the energy per unit length needed to break the cohesive bonds. Therefore, it is simply the bare surface energy $\gamma$
\begin{equation}
\gamma\equiv\frac{1}{2}(1-u_{-}^2) \ .
\end{equation}
The second term on the left-hand-side is the frictional/viscous dissipation. Together, these two terms result in the fracture energy $\Gamma(v)$ \cite{98Fre}
\begin{equation}
\Gamma(v,\eta)=\gamma+\eta v \int_{-\infty}^0(\pa_xu)^2dx \ .
\end{equation}
The right-hand-side of Eq. (\ref{energy_balance}) is the energy per unit length stored far ahead of the crack tip. Therefore, this is the so-called ``energy release rate'' $G(v)$ \cite{98Fre}
\begin{equation}
G(v)=\frac{1}{2}\kappa^2(\delta-u_{-})^2=\frac{\delta^2(v)\kappa^2}{2(1+\kappa^2)^2}\ ,
\end{equation}
where $\delta(v)$ is obtained by inverting Eq. (\ref{vSS}). Therefore, we can rewrite the energy balance of Eq. (\ref{energy_balance}) in the common LEFM form $G(v)=\Gamma(v,\eta)$ \cite{98Fre}. By reexamining Eq. (\ref{energy_balance}) we observe that the velocity $v$ is coupled to $\eta$ such that when $\eta$ vanishes $v$ does not appear in the equation. Therefore, for $\eta\!=\!0$, there exist steady states only when $\delta\!=\!\delta_c$ and in this case, since $G(v)=\Gamma(v,0)$ independently of $v$, the crack can propagate at any velocity. For $\delta\!>\!\delta_c$ we have $G\!>\!\Gamma$ and there exist no steady states. In that case the crack is expected to accelerate toward the limiting velocity $v\!=\!1$. This result is consistent with the $\eta\!\to\!0$ limit of Eq. (\ref{vSS}). However, one should be cautious as this limit predicts that $v\!=\!1$ also for $\delta\!\le\!\delta_c$, which in light of the discussion above, is wrong. This observation was made previously for a related model \cite{93Mar}.

In order to set the stage for the comparison between the linear and nonlinear models below, we should ask what one can learn about this comparison from the steady states of the linear model alone. Intuitively, it is quite clear what determines the range of validity of the linear approximation: it is expected to breakdown if strains significantly larger than a few percent develop over a large enough region near the crack tip. Mathematically speaking, we expect the linear approximation to hold if $(\pa_xu)^2$ is sufficiently smaller than $\pa_xu$ almost everywhere. In order to make this qualitative observation more quantitative, we note that the largest strain derived from Eq. (\ref{solution_linearSS}), which occurs at $x\!=\!0$, is
\begin{equation}
\label{SIF}
\pa_xu(0)=\frac{(\delta-1)\kappa}{\sqrt{1-v^2}} \ .
\end{equation}
This quantity can be interpreted as the analog of the stress intensity factor of LEFM \cite{98Fre}, since it provides a measure of the typical strains near the tip of the crack. Moreover, it is an increasing function of the load $\delta$ and $\kappa$, and more importantly, it {\em diverges} in the limit $v\!\to\!1$. Thus, we already learn that even if there exists a low velocities range where the linear approximation holds, then it will breakdown at a finite, possibly high, velocity. In fact, the issue of whether there exists a low velocities range of validity of the linear approximation can be further elucidated by substituting $\delta\!=\!\delta_c$ and $v\!=\!0$ in Eq. (\ref{SIF}) to obtain the following inequality
\begin{equation}
\label{bound}
\pa_xu(x;v\!=\!0)\le(\sqrt{1+\kappa^{-2}}-1)\kappa\le\pa_xu(x\!=\!0;v) \ .
\end{equation}
This inequality suggests that the value of $(\sqrt{1+\kappa^{-2}}-1)\kappa$, in comparison to unity, determines whether there exists a range of low velocities where the linear approximation is valid or not. For example, for $\kappa\!=\!2$ we obtain $(\sqrt{1+\kappa^{-2}}-1)\kappa\!\simeq\!0.24$ and we expect the linearized model to provide a reasonable approximation to the nonlinear model at low velocities (see Fig. \ref{nonlin}), while for $\kappa\!=\!0.3$ we obtain $(\sqrt{1+\kappa^{-2}}-1)\kappa\!\simeq\!0.74$, for which we expect that no range of validity for the linear approximation exists.
In the next subsection we solve for the steady states of the nonlinear models and present a detailed comparison with the results obtained above for the linear model.

\subsection{The nonlinear models}
\label{SSnonlin}

%%%%%%% FIGURE 3 %%%%%%%%%%%%%%%%%%
\begin{figure}
\centering
\epsfig{width=.4\textwidth,file=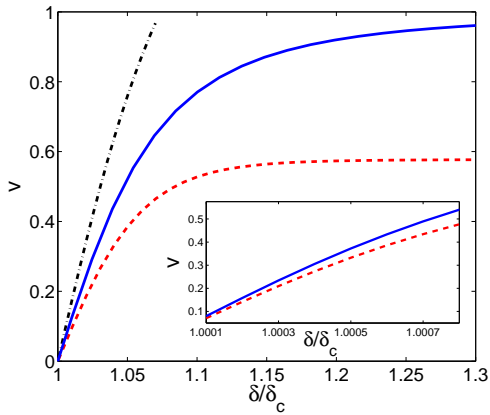}
\caption{(Color online) Main panel: The velocity $v$ vs. the normalized load $\delta/\delta_c$ for the linear model (solid line) - Eq. (\ref{vSS}), the nonlinear softening model (dashed line) and the nonlinear stiffening model (dashed-dotted line), all with $\kappa\!=\!0.3$ and $\eta\!=\!0.3$. The result for the linear model is given in Eq. (\ref{vSS}). Note that there exist no smooth steady states with $v\!>\!1$ in the nonlinear stiffening model. In this range shock behavior is expected (not shown here).
Inset: The velocity $v$ vs. the normalized load $\delta/\delta_c$ for the linear model (solid line) and the nonlinear softening model (dashed line) with $\kappa\!=\!2$ and $\eta\!=\!0.1$.}\label{Load_Velocity}
\end{figure}
%%%%%%%%%%%%%%%%%%%%%%%%%%%%%%%%%%%

Our aim in this subsection is to solve for the steady states of the nonlinear models and compare the results with those of the linear model.
Substituting Eq. (\ref{NNF}) into Eq. (\ref{nondim}), using the steady state assumption $u(x\!\pm\! vt)$ and replacing $x'\!=\!x\pm\! vt$ with $x$ for the simplicity of the notation, we obtain
\begin{eqnarray}
\label{EOM_nonlinearSS}
\left(\frac{1}{3}-v^2\right)\pa_{xx}u\!&+&\!\frac{2\pa_{xx}u}{3(1+\pa_xu)^3}=\\
\kappa^2(u-\delta)\!&+&\!\left(u\mp v\eta\pa_xu\right)H(\mp x)\ .\nonumber
\end{eqnarray}
We note that the $\mp$ signs correspond to nonlinear softening and nonlinear stiffening respectively. For the former the crack propagates in the negative x-direction and $\pa_x u\!>\!0$ such that the softening branch of the constitutive law of Eq. (\ref{NH}) is selected (see Fig. \ref{nonlin}), while for the latter the crack propagates in the positive x-direction and $\pa_x u\!<\!0$ such that the stiffening branch of the constitutive law of Eq. (\ref{NH}) is selected (see Fig. \ref{nonlin}).
The appearance of nonlinearities in Eq. (\ref{EOM_nonlinearSS}) entails a numerical solution. We solve the problem by using the shooting method. We first guess $\pa_xu(0)$ and $v$, and then integrate Eq. (\ref{EOM_nonlinearSS}) from $x\!=\!0$ in both directions using a forth order Runge-Kutta integration scheme, where $u(0)\!=\!1$ is used. We then improve the guess until $u(x\!\to\!-\infty)\!\to\!\delta$ and $u(x\!\to\!\infty)\!\to\!\delta \kappa^2/(1+\kappa^2)$ are approached monotonically.

We first consider the velocity-load relations for both the linear and nonlinear models. In the main panel of Fig. \ref{Load_Velocity} the propagation velocity $v$ is plotted as a function of $\delta/\delta_c$ for the linear and nonlinear models with $\kappa\!=\!0.3$ and $\eta\!=\!0.3$. We first discuss the effect of elastic nonlinearities on the limiting crack velocity. In the linear case (solid line), as discussed in relation to Eq. (\ref{vSS}), the limiting velocity is $v\!\to\!1$, which is the linear elastic wave speed. In the nonlinear softening case (dashed line) the limiting velocity is substantially smaller, $v\!\to\!1/\sqrt{3}$. To understand this, recall that the limiting crack velocity is determined by the speed of small amplitude waves traveling near the tip of the crack, since these waves determine the rate at which energy is being transferred to the crack tip for breaking cohesive bonds. The speed of these waves, in higher dimensional models, is determined by the properties of the bulk material. In the simple one-dimensional models considered here there is no clear separation between the ``bulk'', that is represented by the elastic membrane, the external loading that is represented by the
$\kappa^2(u-\delta)$ term and the cohesive force $\phi(u,\pa_t u)$; therefore, the speed of small amplitude waves is affected by the loading and interfacial cohesion, in addition to the ``bulk'' properties. However, for the sake of obtaining a physical understanding of the effect of elastic nonlinearities on the limiting crack velocity we consider in the discussion below only the elastic properties of the membrane.

In the linear model the small amplitude wave speed is independent of deformation and equals to the small strains wave speed, therefore $v\!\to\!1$ in our dimensionless units. However, in the nonlinear models the small amplitude wave speed, that is determined by the local tangent to the stress-strain curve, {\em depends} on the state of deformation near the moving crack tip. In the softening branch of the stress-strain curve presented in Fig. \ref{nonlin}, the local tangent decreases with increasing strain until it approaches $1/3$. Since the crack tip concentrates large strains, and the magnitude of these strains increases with increasing propagation velocities, the tip velocity is determined by the square root of the limiting local tangent $1/\sqrt{3}$. In the stiffening branch of the stress-strain curve presented in Fig. \ref{nonlin}, the local tangent continuously increases with increasing the magnitude of the strain (the strain itself is negative in this case). Therefore, we expect cracks to propagate faster in the nonlinear stiffening model, compared to the linear model. Moreover, if the crack propagates at velocities {\em higher} than the small strains wave speed, we expect the development of shocks. The velocity-load relation for the nonlinear stiffening model is shown in Fig. \ref{Load_Velocity} (dashed-dotted line). As expected, for a given load, the propagation velocity is higher than in the linear and nonlinear softening models. Moreover, when the velocity approaches the small strains wave speed $v\!\to\!1$, we failed to find {\em smooth} steady states. This point marks the onset of shock development. Note, however, that we expect the existence of steady states with $v\!>\!1$, thus the limiting velocity is expected to be higher than in the linear case. Similar ideas about the effect of elastic nonlinearities on the limiting crack velocity were discussed previously in the literature \cite{96Gao,03BAG,06BG}.

%%%%%%% FIGURE 4 %%%%%%%%%%%%%%%%%%
\begin{figure}
\centering
\epsfig{width=.48\textwidth,file=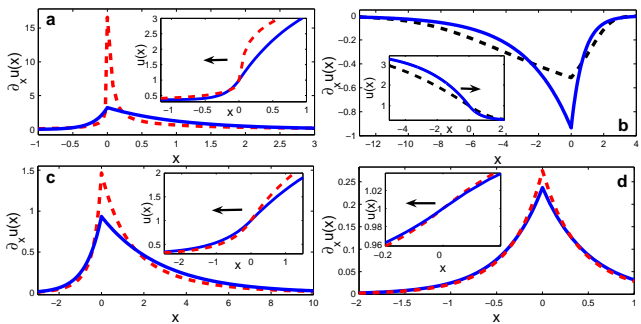}
\caption{(Color online) (a) Main panel: The steady state strain $\pa_xu(x)$ for the linear (solid line) and nonlinear softening (dashed line) models with $\delta\!=\!1.25\delta_c$, $\kappa\!=\!0.3$ and $\eta\!=\!0.3$. Note that the velocities of the two profiles are different, see Fig. \ref{Load_Velocity}. Inset: The displacement $u(x)$ in the near crack tip region. The arrow indicates the direction of propagation. (b) The same as (a), but for the linear (solid line) and nonlinear stiffening (dashed line) models with $\delta\!=\!1.05\delta_c$.
(c) The same as (a), but with $\delta\!=\!1.05\delta_c$. (d) The same as (a), but with $\delta\!=\!1.0001\delta_c$, $\kappa\!=\!2$ and $\eta\!=\!0.1$}\label{Profiles}
\end{figure}
%%%%%%%%%%%%%%%%%%%%%%%%%%%%%%%%%%%

As discussed at the end of the previous subsection, we expect the linear approximation to fail at all velocities for $\kappa\!=\!0.3$. This is demonstrated in Fig. \ref{Load_Velocity} by the large difference in propagation velocities. In fact, even in the limit $\delta/\delta_c\!\to\!1$ the differences in the velocity of propagation are large due to the significantly different initial slopes of the velocity-load curves. In the inset of Fig. \ref{Load_Velocity} we compare the velocity-load relations for the linear and nonlinear softening models with $\kappa\!=\!2$ and $\eta\!=\!0.1$. For this value of $\kappa$ we expect the linear model to provide a reasonable approximation to the nonlinear model {\em at small velocities}. Indeed, the inset of Fig. \ref{Load_Velocity} shows that at least as far as the velocity of propagation is concerned the velocities in the low $v$ regime are similar in the two models, though they separate progressively with increasing load.

We now turn to compare the deformation in the various models. In the main panel of Fig. \ref{Profiles}a the strain distributions for the linear and nonlinear softening models with $\delta\!=\!1.25\delta_c$, $\kappa\!=\!0.3$ and $\eta\!=\!0.3$ are shown. In light of the results shown in Fig. \ref{Load_Velocity}, we expect large differences in the strain near the crack tip for such a value of the load. Indeed, rather dramatic differences are observed near the tip of the crack, where the strain in the nonlinear softening case is significantly higher than in the linear case. This effect is in qualitative agreement with the findings of \cite{08LBF,08BLF}. In the inset, the displacement in the near tip region is shown. We note that according to Eq. (\ref{energy_balance}) the difference in the integral of $(\pa_xu)^2$ over the negative x-axis determines the difference in propagation velocity when $\delta$, $\kappa$ and $\eta$ are fixed. The large difference in the near tip strains observed in Fig. \ref{Profiles}a, is consistent with the large difference in velocity for $\delta\!=\!1.25\delta_c$ in Fig. \ref{Load_Velocity}. In Fig. \ref{Profiles}b the strain and displacement fields
for the linear and nonlinear stiffening models with $\delta\!=\!1.05\delta_c$, $\kappa\!=\!0.3$ and $\eta\!=\!0.3$ are shown. In this case the nonlinear strains are smaller than the linear ones. The corresponding comparison (i.e. for $\delta\!=\!1.05\delta_c$) with the nonlinear softening model is shown in \ref{Profiles}c. Figs. \ref{Profiles}b and \ref{Profiles}c, both with $\delta\!=\!1.05\delta_c$ exhibit smaller differences between the linear and nonlinear models compared to Fig. \ref{Profiles}a where $\delta\!=\!1.25\delta_c$, demonstrating quantitatively how the linear approximation deteriorates with increasing crack velocity. Note, however, that even for $\delta\!=\!1.05\delta_c$ the differences are non-negligible, in agreement with Fig. \ref{Load_Velocity} that implies that there is no range of validity for the linear approximation for $\kappa\!=\!0.3$. In Fig. \ref{Profiles}d the strain and displacement fields
for the linear and nonlinear softening models with $\delta\!=\!1.0001\delta_c$, $\kappa\!=\!2$ and $\eta\!=\!0.1$ are shown, cf. the inset of Fig. \ref{Load_Velocity}. For this larger value of $\kappa$, the linear model provides a reasonable approximation for the nonlinear one, where moderate differences in strain are observed only very near vicinity of the crack tip.

\begin{figure}
\centering
\epsfig{width=.5\textwidth,file=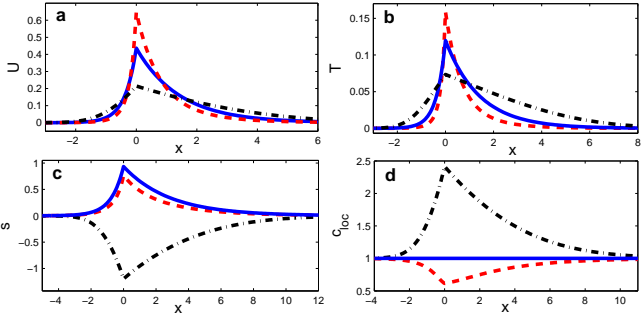}
\caption{(Color online) (a) The steady state potential energy distribution $U$ for the linear (solid line), nonlinear softening (dashed line) and nonlinear stiffening (dashed-dotted line) models with $\delta\!=\!1.05\delta_c$, $\kappa\!=\!0.3$ and $\eta\!=\!0.3$. (b) The same as (a), but for the steady state kinetic energy distribution $T$. (c) The same as (a), but for the steady state stress distribution $s$. (d) The same as (a), but for the small amplitude local wave speed distribution $c_{loc}$ of Eq. (\ref{cNL}). In all panels the data for the nonlinear stiffening model was transformed according to $x\!\to\!-x$ for the sake of comparison with the other models.}\label{energy}
\end{figure}

In Fig. \ref{energy}a the elastic strain energy distributions for the three models with $\delta\!=\!1.05\delta_c$, $\kappa\!=\!0.3$ and $\eta\!=\!0.3$ are shown. The strain energy functional corresponding to the stress-strain relation of Eq. (\ref{NH}) is
\begin{equation}
\label{strain_energy}
U=\frac{1}{3}\left[\frac{1}{2}(1+\pa_xu)^2+\frac{1}{1+\pa_xu}-\frac{3}{2} \right] \ ,
\end{equation}
where Eq. (\ref{NH}) is recovered using the following functional derivative
\begin{equation}
s=\frac{\delta U}{\delta (\pa_xu)} \ .
\end{equation}
Note that in the common linear elastic approximation, Eq. (\ref{strain_energy}) yields
\begin{equation}
\label{strain_energy_LE}
U\simeq\frac{1}{2}(\pa_xu)^2+{\C O}\left[(\pa_xu)^3\right]\ .
\end{equation}
In Fig. \ref{energy}b the kinetic energy distributions $T\!=\!v^2(\pa_xu)^2/2$ for the three models with $\delta\!=\!1.05\delta_c$, $\kappa\!=\!0.3$ and $\eta\!=\!0.3$ are shown. The corresponding stress distributions are shown in Fig. \ref{energy}c. Note that the magnitude of the stress at the tip is similar for the linear and nonlinear softening models, but is different for the nonlinear stiffening case. This can be understood as follows: for the linear model we have $s\!\simeq\!\pa_xu$. For the nonlinear softening model Eq. (\ref{NH}) is approximated as $s\!\simeq\!\case{1}{3}(1+\pa_xu)$ for the large strains near the tip. The difference between these expressions is compensated by the larger strains in the nonlinear softening case, cf. Fig. \ref{Profiles}c, yielding similar values for $s$. However, for the nonlinear stiffening model, Eq. (\ref{NH}) is approximated as $s\!\simeq\!-\case{1}{3}(1+\pa_xu)^{-2}$ for the large negative strains near the tip, cf. Fig. \ref{Profiles}b, resulting in a negative and somewhat larger in magnitude stress in this case. In spite of the fact that the details of the field distributions appearing in Figs. \ref{energy}a-\ref{energy}c are specific to the simple models considered here, they give one a sense of the type of errors expected in models that exclude elastic nonlinearities even at moderate crack propagation velocities.
Possible implications of these differences for questions of crack tip stability will be discussed in Sect. \ref{stability}.
%%%%%%% FIGURE 5 %%%%%%%%%%%%%%%%%%
%%%%%%%%%%%%%%%%%%%%%%%%%%%%%%%%%%%

The last issue to be discussed in relation to the steady state solutions concerns material lengthscales. It is well known that linear elasticity contains no intrinsic lengthscale, while the instabilities of dynamics fracture indicate that some non-geometrical lengthscale is involved \cite{07LBDF, 96SF,05BP,07BP}. The existence of elastic nonlinearities naturally suggests a lengthscale \cite{08LBF,08BLF}, which is simply the size of the nonlinear zone. More precisely, a lengthscale can be defined as the size of the region in which material properties become {\em deformation dependent}, for example the region where the small amplitude local wave speed differs from the linear elastic wave speed. Therefore, we define the small amplitude local wave speed $c_{loc}(x)$ as
\begin{equation}
\label{cNL}
c_{loc}(x)=\sqrt{\frac{\delta s\left[u(x)\right]}{\delta (\pa_xu)}} \ .
\end{equation}
This quantity is plotted in Fig. \ref{energy}d for the three models, demonstrating the appearance of a dynamical lengthscale associated with the nonlinear elastic zone in the crack tip region, a lengthscale that is absent in the linear model. It is a {\em dynamical} lengthscale in the sense that it emerges as a result of the dynamics of the crack. One possible implication of such a lengthscale was discussed in \cite{03BAG, 06BG}, while additional possibilities should be further investigated.

In summary, in this section we have presented a detailed comparison of the steady state solutions of the linear and nonlinear models. We demonstrated that the linear approximation breaks down inevitably at sufficiently large velocity, if there exists a range of validity for that approximation at all. We showed that the linear and nonlinear models differ in their limiting velocities, their near tip strain, stress and energy distributions and in the emergence of a dynamical lengthscale associated with a nonlinear elastic zone. In the next section we focus on the response of the cracks in these models to small perturbations out of steady state.

\section{Response to perturbations: linear stability analysis}
\label{stability}

The results presented up now indicate that the linear approximation breaks down inevitably at sufficiently high velocities. All these results were restricted to steady state conditions. However, as mentioned above, one of the great theoretical challenges in the field of fracture mechanics is the understanding of the origin of crack tip instabilities \cite{99MF,07LBDF}. In this particular respect, the one-dimensional model is certainly too simple as we do not expect any instabilities to occur here. More specifically, since the crack in the one-dimensional model is, by dimensionality alone, restricted to follow a straight path, it can at most change its velocity along this predetermined path. However, it cannot accelerate or decelerate significantly due to the global energy balance constraint. In contradistinction, the tip instabilities observed experimentally \cite{99MF,07LBDF} involve in an essential way the deviation of the crack from the pre-instability straight path. Bearing this limitation of the one-dimensional models in mind and expecting no instability in this framework \cite{93Mar,95CLN}, we still want to study the response of the steady state cracks in these simple models to small perturbations. The motivation for that is to gain some insights (or hints) about the kind of near crack tip physics that is overlooked by the common exclusion of elastic nonlinearities, especially as far as perturbations are considered.

For that aim we perform a linear stability analysis for both the linear and nonlinear models. We stress again that we do not expect any instability to occur, but rather we are interested in the effect of nonlinearities on the relaxation time back to the stable steady state. We start by defining the coordinate transformation $x'\!=\!x\!-\!x_{tip}(t)$ and $t'\!=\!t$, where $(x,t)$ is a fixed coordinate system and $(x',t')$ is a coordinate system that moves with the crack tip $x_{tip}(t)$. As discussed in \cite{93Mar,95CLN}, such a transformation is essential in order to avoid irregular behavior at the crack tip. Defining $\hat{u}(x'(x,t),t')\!\equiv\!u(x,t)$ and using Eqs. (\ref{nondim}) and (\ref{LF}), we obtain for the linear model
\begin{eqnarray}
\label{Tip_coordinate}
\pa_{tt}\hat{u}&\!\!-\!\!&2\dot{x}_{tip}\pa_{xt}\hat{u}+\dot{x}_{tip}^2\pa_{xx}\hat{u}-\ddot{x}_{tip}\pa_x\hat{u}-\pa_{xx}\hat{u}=\\
&\!\!-\!\!&\kappa^2(\hat{u}-\delta)-\left(\hat{u}+\eta\pa_t\hat{u}-\eta\dot{x}_{tip}\pa_x\hat{u}\right)H(1-\hat{u})\ .\nonumber
\end{eqnarray}
Note that we renamed $x'\!\to\!x$ and $t'\!\to\!t$ for notational simplicity.
We are now interested in the time evolution of small perturbations of amplitude $\epsilon$ around the steady state crack tip location
\begin{equation}
\label{perturb}
x_{tip}(t)=-vt-\epsilon e^{\omega t},\quad \hat{u}(x,t)=u(x)-\epsilon \tilde{u}(x) e^{\omega t} \ .
\end{equation}
Substituting these expressions into Eq. (\ref{Tip_coordinate}) and linearizing in $\epsilon$, we obtain
\begin{eqnarray}
\label{EOM_linearPertub}
&\!\!-\!\!&\left(2v\omega\pa_x+\omega^2 \right)\left(\pa_xu-\tilde{u} \right)- (1-v^2)\pa_{xx}\tilde{u}+\kappa^2\tilde{u}=\nonumber\\
&\!\!-\!\!&\left(\tilde{u}+\eta v \pa_x\tilde{u}+\eta \omega \tilde{u} -\eta \omega \pa_xu\right) H(-x)\ .
\end{eqnarray}
Note that $u(x)$ is simply the steady state solution given in Eq. (\ref{solution_linearSS}) and that Eq. (\ref{EOM_linearPertub}) admits a trivial solution with $\tilde{u}\!=\!0$ and $\omega\!=\!0$. This solution corresponds to a translation of the steady state solution and is of no interest here.

Using $\hat{u}(0)\!=\!1$ (recall that $x\!=\!0$ is still the crack tip location) and $u(0)\!=\!1$, we obtain $\tilde{u}(0)\!=\!0$. Furthermore, substituting $\tilde{u}=\pa_xu+\bar{u}$ into Eq. (\ref{EOM_linearPertub}), we obtain a simpler problem for $\bar{u}$, with $\bar{u}(0)\!=\!-\pa_xu(0)$. The resulting problem can be rather easily solved following a similar procedure to the one employed in solving for the steady states. Specifically, by demanding that $\tilde{u}(0)$ is continuous, we obtain
\begin{widetext}
\begin{equation}
\label{omega}
\omega(v,\kappa,\eta) = \frac{-4\eta\kappa^2\left[4(1-v^2)+\eta^2(1+v^2) \right]+2\eta\kappa(4-\eta^2)\sqrt{(1-v^2)\left[\eta^2v^2+4(1+\kappa^2)(1-v^2) \right]}}{(\eta^2-4)^2(v^2-1)+16\eta^2\kappa^2}\ .
\end{equation}
\end{widetext}
In the range of interest, i.e. for small $\eta$, we find that $\omega\!<\!0$, implying linear stability as expected.
This result implies that the perturbation of the tip location relaxes with a typical timescale $\tau\!=\!|\omega|^{-1}$.
In the limit $\kappa,\eta\!\ll\!1$ we obtain $\tau\!\simeq\!(\eta\kappa)^{-1}$, which is independent of $v$.
%%%%%%% FIGURE 6 %%%%%%%%%%%%%%%%%%
\begin{figure}
\centering
\epsfig{width=.45\textwidth,file=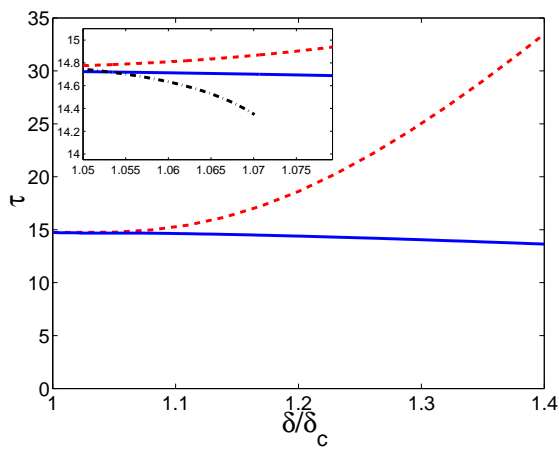}
\caption{(Color online) Main panel: The relaxation time $\tau$ vs. $\delta/\delta_c$ for the linear (solid line) and nonlinear softening (dashed line) models with $\kappa\!=\!0.3$ and $\eta\!=\!0.3$. Inset: Zoom in on the low $\delta/\delta_c$ region, where the relaxation time for nonlinear stiffening  model is added (dashed-dotted line).}\label{tau}
\end{figure}
%%%%%%%%%%%%%%%%%%%%%%%%%%%%%%%%%%%

In order to compare the relaxation time in the linear model to relaxation times in the nonlinear models, we repeat the linear stability analysis for the latter. Using the same notation as before and Eqs. (\ref{nondim})-(\ref{NNF}) with
\begin{equation}
\label{perturb1}
x_{tip}(t)=\pm vt\pm\epsilon e^{\omega_{_{\pm}}\! t},\quad \hat{u}(x,t)=u(x)\pm\epsilon \tilde{u}(x) e^{\omega_{_{\pm}}\! t} \ ,
\end{equation}
we obtain
\begin{eqnarray}
\label{EOM_nonlinearPertub}
&\!\!\mp\!\!&\left(2v\omega_{_{\pm}}\pa_x\pm\omega_{_{\pm}}^2 \right)\left(\pa_xu-\tilde{u} \right)+\kappa^2\tilde{u} =\nonumber\\
&\!\!-\!\!&\left(v^2-\frac{1}{3}\right)\pa_{xx}\tilde{u}+\frac{2\pa_{xx}\tilde{u}}{3(1+\pa_xu)^3}-\frac{2\pa_x\tilde{u}\pa_{xx}u}{(1+\pa_xu)^4}\nonumber\\
&\!\!-\!\!&\left(\tilde{u}\pm\eta v \pa_x\tilde{u}+\eta \omega_{_{\pm}} \tilde{u} -\eta \omega_{_{\pm}} \pa_xu\right) H(\mp x)\ .
\end{eqnarray}
This is the counterpart of Eq. (\ref{EOM_linearPertub}). Note that the $\pm$ in Eq. (\ref{perturb1}) correspond to the stiffening and softening models respectively. Eq. (\ref{EOM_nonlinearPertub}) is solved numerically using a method similar to the one used to obtain the steady state solution. The relaxation times are given as before by $|\omega_{_{\pm}}|^{-1}$. The results are summarized in Fig. \ref{tau}, where the relaxation times for the linear and nonlinear models with $\kappa\!=\!0.3$ and $\eta\!=\!0.3$ are plotted as a function of $\delta/\delta_c$. The results presented in the main panel show that the relaxation time in the nonlinear softening model is {\em larger} than the relaxation time in the linear model. The slower dynamics in the nonlinear softening case may be attributed to the smaller local wave speeds in the near crack tip vicinity. In the inset we focus on the small $\delta/\delta_c$ where smooth steady states for the nonlinear stiffening exist, cf. Fig. \ref{Load_Velocity}. The results show that the relaxation time in the nonlinear stiffening model is {\em smaller} than the relaxation time in the linear model. The faster dynamics in the nonlinear stiffening case may be attributed to the larger local wave speeds in the near crack tip vicinity. The main result obtained here is that the typical response timescale near the crack tip is affected by elastic nonlinearities. This timescale may be of prime importance in understanding the experimentally observed crack tip instabilities \cite{99MF, 07LBDF}. Moreover, this dynamical timescale can be interpreted as introducing inertia-like effects into the crack tip dynamics \cite{inertia}, effects that are missing in LEFM \cite{98Fre}.

\section{Concluding remarks}
\label{sum}

In this paper we investigated the role of elastic nonlinearities in simple one-dimensional models of fracture. We were mainly motivated by the recent experimental and theoretical findings of \cite{08LBF,08BLF} that demonstrated explicitly the importance of elastic nonlinearities for understanding the structure of the deformation near a moving crack tip. Our results show that the common linear elastic approximation breaks down at sufficiently high propagation velocities, if there exists a small velocities range of validity at all. This finding is in complete agreement with the results of  \cite{08LBF,08BLF}, where it was shown that at high velocities the linear elastic approximation of LEFM provides un-physical and qualitatively different results compared to the nonlinear theory.

The breakdown of the linear approximation manifests itself in marked differences in the propagation velocities, including the limiting crack velocity, as well as in the stress and strain distributions in the crack tip vicinity. The near tip deformation, that is markedly different from the linear elastic prediction both in our simple model and in Refs. \cite{08LBF,08BLF}, may have a role in determining the stability of the crack tip against perturbations. In this regard, we demonstrated the existence of a lengthscale that is associated with the nonlinear elastic zone surrounding the crack tip. This lengthscale was shown in \cite{08BLF} to coincide with the wavelength of the oscillations observed in \cite{07LBDF}. This finding can potentially explain the emergence of a non-geometrical lengthscale that is missing in the standard approach of LEFM \cite{07BP}. Furthermore, by studying the response of the crack tip to perturbations we showed that elastic nonlinearities affect the crack tip local response timescale. This emerging timescale can be interpreted as effectively attributing inertia-like properties to the crack tip \cite{inertia}, in contradistinction with LEFM where the crack tip is regarded as ``massless'' \cite{98Fre}.

The simple one-dimensional models considered in this work offer some insights about the possible importance of elastic nonlinearities in answering the long standing question of ``how things break?'' However, in order to obtain concrete predictions related to the experimentally observed instabilities, one should study crack propagation with near tip elastic nonlinearities in higher dimensions. The novel asymptotic nonlinear solution presented in \cite{08BLF} may serve as a promising starting point.

{\bf Acknowledgements} E. Bouchbinder acknowledges support from the Horowitz Center for Complexity Science and the Lady Davis Trust.


\begin{thebibliography}{99}

\bibitem{99MF} M. Marder and J. Fineberg, Phys. Rep. {\bf 313}, 1 (1999).

\bibitem{07LBDF} A. Livne, O. Ben-David and J. Fineberg, Phys. Rev. Lett. {\bf 98}, 124301 (2007).

\bibitem{98Fre} L. B. Freund, {\em Dynamic Fracture Mechanics}, (Cambridge University Press, Cambridge, 1998).

\bibitem{08LBF} A. Livne, E. Bouchbinder and J. Fineberg, ``The Breakdown of Linear Elastic Fracture Mechanics near the Tip of a Rapid Crack'', arXiv:0807.4866 (2008).

\bibitem{08BLF} E. Bouchbinder, A. Livne and J. Fineberg, ``Weakly Nonlinear Theory of Dynamic Fracture'', arXiv:0807.4868 (2008).

\bibitem{96SF} E. Sharon and J. Fineberg, Phys. Rev. B {\bf 54}, 7128 (1996).

\bibitem{05BP} E. Bouchbinder and I. Procaccia, Phys. Rev. E, {\bf 72}, 055103(R) (2005).

\bibitem{07BP} E. Bouchbinder and I. Procaccia, Phys. Rev. Lett. {\bf 98}, 124302 (2007).

\bibitem{92Lan} J. S. Langer, Phys. Rev. A {\bf 46}, 3123 (1992).

\bibitem{93Mar} M. Marder, Physica D {\bf 66}, 125 (1993).

\bibitem{95CLN} E. S. C. Ching, J. S. Langer and H. Nakanishi, Phys. Rev. E {\bf 52}, 4414 (1995).

\bibitem{98LL} A. E. Lobkovsky and J. S. Langer, Phys. Rev. E {\bf 58}, 1568 (1998).

\bibitem{96Gao} H. Gao, J. Mech. Phys. Solids, {\bf 44}, 1453 (1996).

\bibitem{03BAG} M. J. Buehler, F. F. Abraham and H. Gao, Nature {\bf 426}, 141 (2003).

\bibitem{06BG} M. J. Buehler and H. Gao, Nature {\bf 439}, 307 (2006).

\bibitem{inertia} E. Sharon, G. Cohen and J. Fineberg, Phys. Rev. Lett. {\bf 88}, 085503 (2002).

\end{thebibliography}
\end{document}